\documentclass[aps,prd,preprint,superscriptaddress]{revtex4}

\usepackage{graphicx}
\usepackage{amsmath}
\usepackage{bm}

\def\ee{\end{eqnarray}}

\def\=:{=\hspace{-.7em}\raisebox{1.1ex}{.}\hspace{.1em}\raisebox{-0.2ex}{.} }

\def\ee{\end{eqnarray}}

\def\=:{=\hspace{-.7em}\raisebox{1.1ex}{.}\hspace{.1em}\raisebox{-0.2ex}{.} }

\newcommand {\beq}{\begin{eqnarray}}
\newcommand {\eeq}{\end{eqnarray}}
\newcommand {\non}{\nonumber\\}

\newcommand {\1}[1]{\frac{1}{#1}}

\newcommand {\del}{\partial}

\newcommand {\tr}{{\rm tr}\,}

\begin{document}


\title{Correspondence between Skyrmions 
in 2+1 and 3+1 Dimensions
}


\author{Muneto Nitta}

\affiliation{
Department of Physics, and Research and Education Center for Natural 
Sciences, Keio University, Hiyoshi 4-1-1, Yokohama, Kanagawa 223-8521, Japan\\
}


\date{\today}
\begin{abstract}

A lump (2D Skyrmion) can be constructed as a sine-Gordon kink 
(1D Skyrmion) inside a domain wall in the massive $O(3)$ sigma model. 
In this paper, we discuss relations between Skyrmions in 2+1 and 3+1 dimensions. 
We first construct a Bogomol'nyi-Prasad-Sommerfield 3D Skyrmion as 
a lump inside a non-Abelian domain wall in an $O(4)$ sigma model 
with a potential term admitting two discrete vacua.
Next, we construct a conventional 3D Skyrmion as 
a baby Skyrmion in a non-Abelian domain wall in the Skyrme model 
with a modified mass term admitting two discrete vacua. 
We also construct a spinning 3D Skyrmion as a Q-lump
in the non-Abelian domain wall.

\end{abstract}

\maketitle

\section{Introduction}

The Skyrme model was proposed to describe nucleons as 
topological solitons (Skyrmions) 
characterized by $\pi_3(S^3)\simeq{\bf Z}$ 
in  the pion effective field theory or the chiral Lagrangian \cite{Skyrme:1962vh}. 
Although the nucleons are now known as bound states of quarks, 
the idea of the Skyrme model is still attractive.
In fact, the Skyrme model is still valid as the low-energy description of QCD, 
for instance, in holographic QCD \cite{Sakai:2004cn, Hata:2007mb}.
The Skyrme model can be formulated as an $O(4)$ sigma model 
with a quartic derivative term. 

As discussed by Skyrme himself \cite{Skyrme:1961vr}, sine-Gordon kink 
characterized by $\pi_1(S^1)\simeq{\bf Z}$ 
is a lower-dimensinal toy model of Skyrmion. 
The sine-Gordon model can be formulated as an $O(2)$ sigma model with a potential term,  
and the sine-Gordon kink can be regarded as a one-dimensional (1D) Skyrmion. 
In addition, in two spatial dimensions, 
2D Skyrmions characterized by $\pi_2(S^2)\simeq{\bf Z}$ are known  
\cite{Polyakov:1975yp}, 
which are often called lumps (or sigma model instantons).  
In fact, lumps in an $O(3)$ sigma model with quartic derivative term and a potential term 
are called the baby Skyrmions \cite{Piette:1994ug}.
Therefore, Skyrmions exist in diverse dimensions \cite{Jackson:1988xk}: 
an $O(N+1)$ model model admits $N$ dimensional Skyrmions 
characterized by $\pi_{N}(S^{N})\simeq{\bf Z}$,
at least for $N=1,2,3$. 
Among Skyrmions in diverse dimensions,  
a sine-Gordon kink (1D Skyrmion) was 
obtained as a holonomy of  
a ${\bf C}P^1$ lump (2D Skyrmion) \cite{Sutcliffe:1992ep} 
as a lower-dimensional analog of the Atiyah-Manton construction of 
3D Skyrmion from instanton holonomy \cite{Atiyah:1989dq}.
This relation can be physically explained \cite{Nitta:2012xq} 
with the help of a ${\bf C}P^1$ domain wall \cite{Abraham:1992vb}; 
a sine-Gordon kink in the domain wall theory 
is nothing but a lump in the bulk point of view  
\cite{Kudryavtsev:1997nw,Auzzi:2006ju,Nitta:2012xq}.

In this paper, we discuss the relation between Skyrmions in two and three dimensions.
We first consider an $O(4)$ model with a potential term admitting two discrete vacua. 
This model admits a domain wall solution \cite{Losev:2000mm}. 
With the Skyrme term, the model is the Skyrme model with the modified mass term 
considered before \cite{Kudryavtsev:1999zm},
in which the interaction between the domain wall and the Skyrmions was studied. 
Since the presence of the domain wall solution spontaneously breaks 
$SO(3)$ symmetry of the vacua to $SO(2)$, 
there appear $SO(3)/SO(2)\simeq S^2$ Nambu-Goldstone modes 
in the vicinity of the domain wall. 
Consequently, a continuous family of the domain wall solutions with $S^2$ moduli 
(zero modes , or collective coordinates) exists. 
These $S^2$ zero modes are in fact normalizable and  
the low-energy effective action on the domain wall 
is the $O(3)$ sigma model with 
the target space $S^2$ \cite{Losev:2000mm}. 

Our findings are twofold.
First, in the absence of the Skyrme term in the bulk, 
a lump solution exists in the effective action of the Bogomol'nyi-Prasad-Sommerfield (BPS) 
non-Abelian domain wall.
The lump in the domain wall effective theory 
corresponds to a 3D Skyrmion in the bulk point of view;  
Although 3D Skyrmions in the bulk are unstable to shrink in the absence of  the Skyrme term, 
they can stably exist inside the domain wall as the lumps. 
Since both the domain wall and the lumps are BPS, 
the composite states of 3D Skyrmions are also BPS so that no force exists between 
BPS 3D Skyrmions.  
Second, in the presence of the Skyrme term and the potential term in the bulk, 
we show that the baby Skyrme term \cite{Piette:1994ug} and the potential term are induced 
in the domain wall effective action. 
In this case, the baby Skyrmions \cite{Piette:1994ug} 
in the wall correspond to the conventional 3D Skyrmions in the bulk, 
and 3D Skyrmions can exist both in the bulk and inside the wall.

The baby Skyrme model was proposed as a lower-dimensional toy model of the original Skyrmion.
The result in this paper implies that when 3+1-dimensional Skyrmions are confined on 
a 2+1-dimensional plane, they become baby Skyrmions.
Therefore, the dynamics of baby Skyrmions describe dynamics of 3+1-dimensional Skyrmions confined on a plane.
Since Skyrmions are considered to describe nucleons, 
our result may be applied to a situation that nucleons confined on a plane, 
such as nucleons in the presence of domain walls 
in cores of ferromagnetic neutron stars \cite{Eto:2012qd}.

This paper is organized as follows. 
In Sec.~\ref{sec:model}, we give the Skyrme model with a modified mass term 
and an $O(4)$ sigma model with a potential term, both admitting two discrete vacua.
In Sec.~\ref{sec:NA-wall}, we give a solution of a non-Abelian domain wall 
with the $S^2$ moduli in these theories, 
and construct the effective field theory on the domain wall, 
which is an $O(3)$ sigma model. 
When potential terms are added in the original theory, 
corresponding potential terms are induced in the domain wall effective action. 
Sec.~\ref{sec:lump} is the main part of this paper.
First, we construct sigma model lumps in the domain wall effective theory 
in the theory without the Skyrme term  
and show that they can be identified with 3D Skyrmions in the bulk point of view. 
They are BPS 3D Skyrmions which are unstable in the bulk but 
can stably exist inside the domain wall.
Next, we construct Q-lumps in the wall theory which can be understood as
spinning Skyrmions in the bulk point of view.
Finally, we show that the Skyrme term in the bulk induces the baby Skyrme term 
in the domain wall effective theory, 
so that baby Skyrmions in the wall theory can be identified with 
conventional 3D Skyrmions in the bulk. 
Sec.~\ref{sec:summary}  is devoted to a summary and discussion.

\newpage 
\section{The $O(4)$ sigma model and the Skyrme model\label{sec:model}}
We consider the $SU(2)$ principal chiral model or the Skyrme model 
in $d=3+1$ dimensions. 
With the $SU(2)$ valued field $U(x) \in SU(2)$,
the Lagrangian which we consider is given by
\beq
 {\cal L} = {f_{\pi}^2\over 16} \tr (\del_{\mu}U^{\dagger} \del^{\mu} U) + {\cal L}^{(4)}
 - V(U)
\eeq
with the Skyrme term 
\beq  
 {\cal L}^{(4)}(U) = {1\over 32 e^2} \tr ([U^\dagger \del_{\mu} U, U^\dagger \del_{\nu} U]^2) . 
\label{eq:Skyrme-term}
\eeq
We consider the potential term given by
\beq
 V(U) = m_\pi^2 \tr [(U + U^\dagger -2{\bf 1}_2)(U + U^\dagger +2{\bf 1}_2)]
 \label{eq:modified}
\eeq
which admits two discrete vacua $U = \pm {\bf 1}_2$, 
instead of the conventional potential $V = m_{\pi}^2 [(U + U^\dagger -2{\bf 1}_2)$ 
admitting the unique vacuum $U={\bf 1}_2$. 
The potential term in Eq.~(\ref{eq:modified}) was called a modified mass \cite{Kudryavtsev:1999zm}.

Introducing a four vector of scalar fields $n_i(x)$ $(i=1,2,3,4)$ by
$U(x) = n_4 (x) {\bf 1}_2 + i \sum_{i=1}^3 n_i(x) \sigma_i$ 
with the Pauli matrices $\vec{\sigma}$, 
the Lagrangian can be rewritten in the form of the $O(4)$ model:
\beq
 {\cal L} = \1{2} \partial_{\mu}{\bf n} \cdot \partial^{\mu}{\bf n} 
+ {\cal L}^{(4)}({\bf n})- V({\bf n}) , 
 \quad {\bf n}^2 =1,
\eeq 
with the potential and the Skyrme term, rewritten as
\beq 
V({\bf n}) &=& m^2(1-n_4^2) ,  \label{eq:potential-n} \\
{\cal L}^{(4)} ({\bf n})
&=&   \1{2} (\partial_{\mu}{\bf n} \cdot \partial_{\nu}{\bf n})
             (\partial^{\mu}{\bf n} \cdot \partial^{\nu}{\bf n})
       - \1{2} (\partial_{\mu}{\bf n} \cdot \partial^{\mu}{\bf n} )^2 ,
\eeq
respectively. 
Here we have rescaled the fields and the coordinates. 
The vacua are given by $n_4=\pm 1$. 
The symmetry of vacua is $SO(3)$. 
We work in the field ${\bf n}(x)$ of the $O(4)$ model rather than 
the $SU(2)$ valued field $U(x)$.
\begin{figure}[ht]
\begin{center}
\includegraphics[width=0.6\linewidth,keepaspectratio]{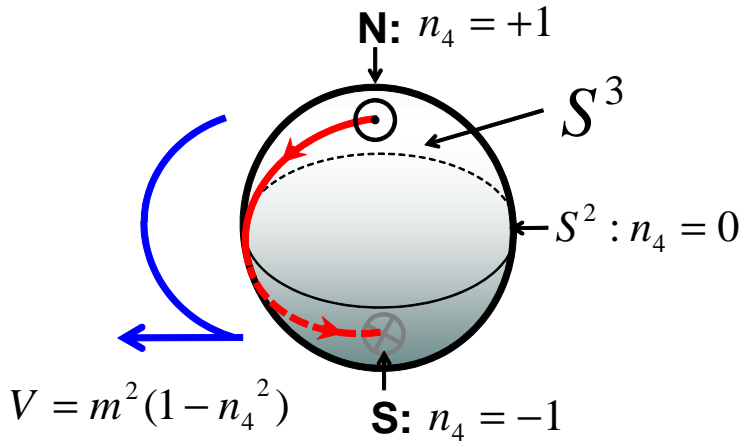}
\caption{The target space $S^3$ of an $O(4)$ sigma model 
with a potential term admitting two discrete vacua. 
The constraint $\sum_{i=1}^4 (n_i(x))^2=1$ defines the $S^3$ target space.
The potential term $V=m(1-n_4^2)$ explicitly breaking $O(4)$ symmetry 
admits the two discrete vacua $n_4 =\pm 1$, denoted by N (north pole) and 
S (south pole).  Each slice at $n_4$= const. is isomorphic to $S^2$ 
except for $n_4=\pm 1$. 
The potential admits a non-Abelian 
domain wall solution with $S^2$ internal moduli, denoted by a curve 
connecting $n_4=\pm 1$.
\label{fig:3sphere}
}
\end{center}
\end{figure}

We later consider the further potential terms which 
explicitly break the $SO(3)$ symmetry into its subgroup $SO(2)$:
\beq 
 V_{\rm linear} (n_{\hat i}) = -m_{\hat i}^2 n_{\hat i}, \quad 
 V_{\rm quad} (n_{\hat i}) = -m_{\hat i}^2 n_{\hat i}^2,  
\quad (\hat i = 1,2,3). 
\label{eq:potential-n2}
\eeq
The former is the conventional mass term for the Skyrme model 
while the latter 
is the same form with Eq.~(\ref{eq:potential-n}). 


\section{Non-Abelian domain wall}\label{sec:NA-wall}

In this section, we construct a non-Abelian domain wall solution 
and the effective field theory of the $S^2$ moduli on the domain wall. 
The domain wall solution was studied in the Skyrme model 
with the modified mass \cite{Kudryavtsev:1999zm} 
and in the $O(4)$ sigma model without the Skyrme term \cite{Losev:2000mm}. 
Here, we work without the Skyrme term 
since the solution is not modified \cite{Kudryavtsev:1999zm}.

\subsection{Non-Abelian domain wall solution}

Let us construct
a domain wall perpendicular to the $x^3$-coordinate, 
connecting the two discrete vacua $n_4 = \pm 1$. 
The energy density of the theory is 
\beq
{\cal E} 
 = \1{2} \partial_{3}{\bf n} \cdot \partial_{3}{\bf n} 
 + m^2(1-n_4^2).
\eeq
By using the $SO(3)$ symmetry acting on $(n_1,n_2,n_3)$, we can put $n_2=n_3 =0$ consistently.  
Using the parametrization ${\bf n} = (\sin \theta, 0, 0, \cos \theta)$,  
the energy density reduces to 
\beq
{\cal E} 
 &=& \1{2} (\partial_{3} n_1)^2 + (\partial_{3} n_4)^2 
 + m^2(1-n_4^2)  \non
 &=& \1{2} (\partial_3 \theta)^2 + m^2 \sin^2 \theta.
\eeq
This is the sine-Gordon model. 
The Bogomol'nyi completion for the domain wall can be obtained as
\beq
E &=& \1{2} \int d x^3 
[(\partial_3 \theta \mp \sqrt 2 m \sin \theta)^2 
   \pm 2\sqrt 2 m \partial_3 \theta \sin \theta ]
\nonumber \\ 
 &\geq& |T_{\rm wall}| , \label{eq:BPS-bound-wall}
\eeq
where, $T_{\rm wall}$ is 
the topological charge that characterizes the wall:
\begin{eqnarray}
T_{\rm wall} = \pm \sqrt 2 m \int  d x^3 (\partial_3 \theta \sin \theta) 
 = \mp \sqrt 2 m [\cos \theta]^{x^3 =+\infty}_{x^3 =-\infty}  
 = \mp \sqrt 2 m [n_4]^{x^3 =+\infty}_{x^3 =-\infty}  = 2 \sqrt 2 m  \label{eq:tension} .
\end{eqnarray}
Among all configurations with a fixed boundary condition, 
that is, with a fixed topological charge $T_{\rm wall}$, 
the most stable configurations with 
the least energy saturate the inequality (\ref{eq:BPS-bound-wall}) 
and satisfy the BPS equation  
\begin{equation}
\partial_3 \theta \mp \sqrt 2 m \sin \theta=0,  \label{eq:BPSeq}
\end{equation}
which is obtained by $|...|^2=0$ in  Eq.~(\ref{eq:BPS-bound-wall}). 
This BPS equation can be immediately solved as 
\beq
 \theta(x^3) = \arctan \exp (\pm \sqrt 2 m (x^3 -X))
 , \label{eq:wall}
\eeq
with the width $\Delta x^3 = 1/m$, 
where $\pm$ denotes a domain wall and an anti-domain wall.
Here $X$ is a real constant corresponding to the position of the wall.

The most general solution can be obtained by acting 
the vacuum symmetry $SO(3)$ on this particular solution.  
We thus obtain a continuous family of solutions
\beq
&&  n_{\hat i} = \hat n_{\hat i} \sin \theta(x) , ({\hat i}=1,2,3), \quad 
\hat {\bf n}^2 = \sum_{\hat i =1}^3 \hat n_{\hat i}^2=1, \non
&& n_4 = \cos \theta(x)
\eeq
with $\theta$ in Eq.~(\ref{eq:wall}). 
The solution has the moduli parameters 
or the collective coordinates $\hat n_i$ representing $S^2$ in addition to $X$.
Since the presence of the domain wall solution spontaneously breaks the vacuum symmetry $SO(3)$ into its subgroup $SO(2)$, 
the internal orientational moduli $S^2$ correspond to 
the Nambu-Goldstone modes arising from this symmetry breaking. 
Such $S^2$ moduli were found in 
\cite{Losev:2000mm} 
while they were not mentioned in \cite{Kudryavtsev:1999zm}.
Since the symmetry breaking occurs in the vicinity of the domain wall, 
we expect that these modes are normalizable, 
which we shall explicitly demonstrate in the next subsection. 

We call this solution a non-Abelian domain wall in the sense that 
it carries non-Abelian Nambu-Goldstone modes as the moduli parameters.
There is another example of a non-Abelian domain wall 
in a non-Abelian $U(2)$ gauge theory \cite{Shifman:2003uh,Eto:2005cc}, 
which has $SU(2) \simeq S^3$ moduli.
Later, it was generalized to a non-Abelian domain wall with 
$SU(N)$ moduli in $U(N)$ gauge theory \cite{Eto:2008dm}.

\subsection{Low-energy effective theory on domain wall world-volume}
Next, let us construct the effective field theory of the domain wall 
($+$ signature in Eq.~(\ref{eq:wall})). 
According to Manton \cite{Manton:1981mp,Eto:2006uw}, 
the effective theory on the domain wall can be obtained 
by promoting the moduli parameters to fields $X(x^a)$ and $\hat n(x^a)$ 
on the domain wall world-volume $x^a$ ($a=0,1,2$), 
and by performing the integration over the codimension $x\equiv x^3$:
\beq
&& {\cal L}_{\rm dw.eff.} \non
&=& \int_{-\infty}^{+\infty} dx \left[ 
    \1{2} (\partial_{a}n_{\hat{i}} )^2 
 + \1{2} (\partial_{a}n_4)^2
 - \1{2} (\partial_{3}n_{\hat{i}})^2
 - \1{2} (\partial_3 n_4 )^2- m^2 (1-n_4^2)\right] \non
&=& \int_{-\infty}^{+\infty} dx \left[ 
   \1{2} \{ \partial_{a}(\sin \theta \hat {\bf n}) \}^2 
 + \1{2} (\partial_{a}\cos\theta)^2
- \1{2} \left[\partial_{3}(\sin \theta \hat {\bf n}) \right]^2
- \1{2} (\partial_3 \cos\theta )^2- m^2 (1-\cos^2\theta)\right] \non
 &=&  \left(\int_{-\infty}^{+\infty} dx 
\sin^2  \theta\right)\left[ \1{2} (\del_a \hat {\bf n})^2  +  m^2 (\del_a X)^2- 2 m^2 \right]
\eeq
where we have used the BPS equation (\ref{eq:BPSeq}) 
and $\hat {\bf n}^2=1$.  Performing the integration, we reach at
\beq
 {\cal L}_{\rm dw.eff.}  =  { \sqrt 2 \over 2m}  (\del_a \hat {\bf n})^2  
 +{T_{\rm wall} \over 2} (\del_a X)^2 -T_{\rm wall} , \quad 
\hat {\bf n}^2=1  \label{eq:wall-eff}
\eeq
where the constant term coincides with minus the tension $T_{\rm wall}$ of the domain wall 
given in Eq.~(\ref{eq:tension}). 
We thus have shown that the moduli ${\bf R} \times S^2$ are normalizable.
The effective theory of the $S^2$ moduli  $\hat {\bf n}$
is the $O(3)$ model with the target space 
$S^2$ as we expected, while 
the effective theory of the translational modulus $X$ is consistent with 
the Nambu-Goto action 
\beq
 S_{\rm NG} = -T_{\rm wall} \sqrt{1-(\del_a X)^2} 
\eeq
at this order.

In the integration of Eq.~(\ref{eq:wall-eff}), 
we have used the second of the following formulas of the integration: 
\beq
 \int_{-\infty}^{+\infty} dx \sin \theta = {\sqrt 2 \pi \over 2m}, \quad
 \int_{-\infty}^{+\infty} dx \sin^2 \theta = {\sqrt 2 \over m}, \quad 
 \int_{-\infty}^{+\infty} dx \sin^4 \theta = {2\sqrt 2 \over 3 m}, \quad
 \label{eq:formulas}
\eeq
with $\theta$ given in Eq.~(\ref{eq:wall}). 
The rests of the formulas will be used below.

For later use, we also consider the effects of the additional potential 
term (\ref{eq:potential-n2}) 
in the original Lagrangian which explicitly breaks the $SO(3)$ symmetry 
to an $SO(2)$ symmetry. 
We treat them as perturbations with a small mass $m_{\hat i} \ll m$ 
($\hat i =1,2,3$)
and assume the wall solution is not modified significantly. 
Within this approximation, we find the following potential 
terms are induced on the domain wall effective theory: 
\beq 
&& V_{\rm dw.linear} = \int dx V_{\rm linear} (n_{\hat i}) 
= -m_{\hat i}^2 \int dx \sin \theta \hat n_{\hat i}
= - {\sqrt 2 \pi \over 2}  {m_{\hat i}^2 \over m} \hat n_{\hat i} , 
  \label{eq:pot2}\\
&& V_{\rm dw.quad} = \int dx V_{\rm quad}(n_{\hat i}) 
= -m_{\hat i}^2 \int dx (\sin \theta \hat n_{\hat i})^2
= - \sqrt 2  {m_{\hat i}^2 \over m} \hat n_{\hat i}^2 
  \label{eq:pot3}, 
\eeq
where we have used the integration formulas in Eq.~(\ref{eq:formulas}) 
and neglected the constant terms.

\section{Lumps and baby Skyrmions inside the domain wall}\label{sec:lump}
In this section, we construct lumps or baby Skyrmions in the domain wall effective theory 
without or with the Skyrme term in the original Lagrangian, respectively. 

\subsection{BPS 3D Skyrmion inside the wall as lumps: 
Without the Skyrme term}\label{sec:3DSkyrmion-2Dlump} \label{sec:without-Skyrme}
The $O(3)$ model in Eq.~(\ref{eq:wall-eff}) without the potential term 
is known to admit lump or sigma model instanton solutions \cite{Polyakov:1975yp}, 
which are called 2D Skyrmions in condensed matter physics. 
Using the stereographic coordinate $u$ of the Riemann sphere $S^2 \simeq {\bf C}P^1$, 
defined  by
\beq
 u \equiv {\hat n_1 + i \hat n_2 \over 1- \hat n_3},
\eeq 
The domain wall effective Lagrangian given in Eq.~(\ref{eq:wall-eff}) can be rewritten as 
\beq
&& {\cal L} = 
{ \sqrt 2 \over 2m}
 {\partial_{\mu} u^* \partial^{\mu} u \over (1 + |u|^2)^2} .\label{eq:CP1}
\eeq
The energy for static configurations can be written as 
\beq
 {2m \over \sqrt 2}
 E &=& \int d^2 x 
 {\partial_1 u^* \partial_1 u +\partial_2 u^* \partial_2 u 
  \over (1+|u|^2)^2} \non
 &=& \int d^2 x \left[
  {|\partial_1 u \mp i \partial_2 u|^2 \over (1+|u|^2)^2} 
   \pm {i (\partial_1 u^* \partial_2 u - \partial_2 u^* \partial_1 u )
     \over  (1+|u|^2)^2}
  \right] \non
 &\geq& \left| \int d^2 x \,
  {i (\partial_1 u^* \partial_2 u - \partial_2 u^* \partial_1 u )
     \over  (1+|u|^2)^2} 
 \right| = |T_{\rm lump}|.  \label{eq:BPSeq-lump}
\eeq
Therefore, the energy is bound from below by 
the topological charge of the lumps, defined by 
\beq
 T_{\rm lump} 
&\equiv& \int d^2x  {i (\partial_1 u^* \partial_2 u - \partial_2 u^* \partial_1 u )
\over (1+|u|^2)^2}  
= 2 \pi k , \label{eq:lump-charge}
\eeq
with $k \in {\bf Z}$ denoting the lump number.
The inequality is saturated when 
the (anti-) BPS equations for the lumps 
\beq
\partial_1 u \mp i \partial_2 u = 0 \quad
\Longleftrightarrow \quad \bar \del_{\bar z} u =0 
\quad(\mbox{or } \del_z u =0),  \label{eq:BPS-lump-eq}
\eeq
with the complex coordinate $z \equiv x^1 + i x^2$, are satisfied.
The $k$ BPS lump solution is obtained as
\beq
 u_{\rm lump}(z) = \kappa + \sum_{i=1}^k {\lambda_i \over z - z_i}  ,\quad \kappa, z_i, \lambda_i \in {\bf C}, \label{eq:lump-sol}
\eeq
where $z_i$ ($\lambda_i$) correspond to the position (the size and phase) of 
the $i$-th lump.

\begin{figure}[ht]
\begin{center}
\begin{tabular}{cc}
\includegraphics[width=0.5\linewidth,keepaspectratio]{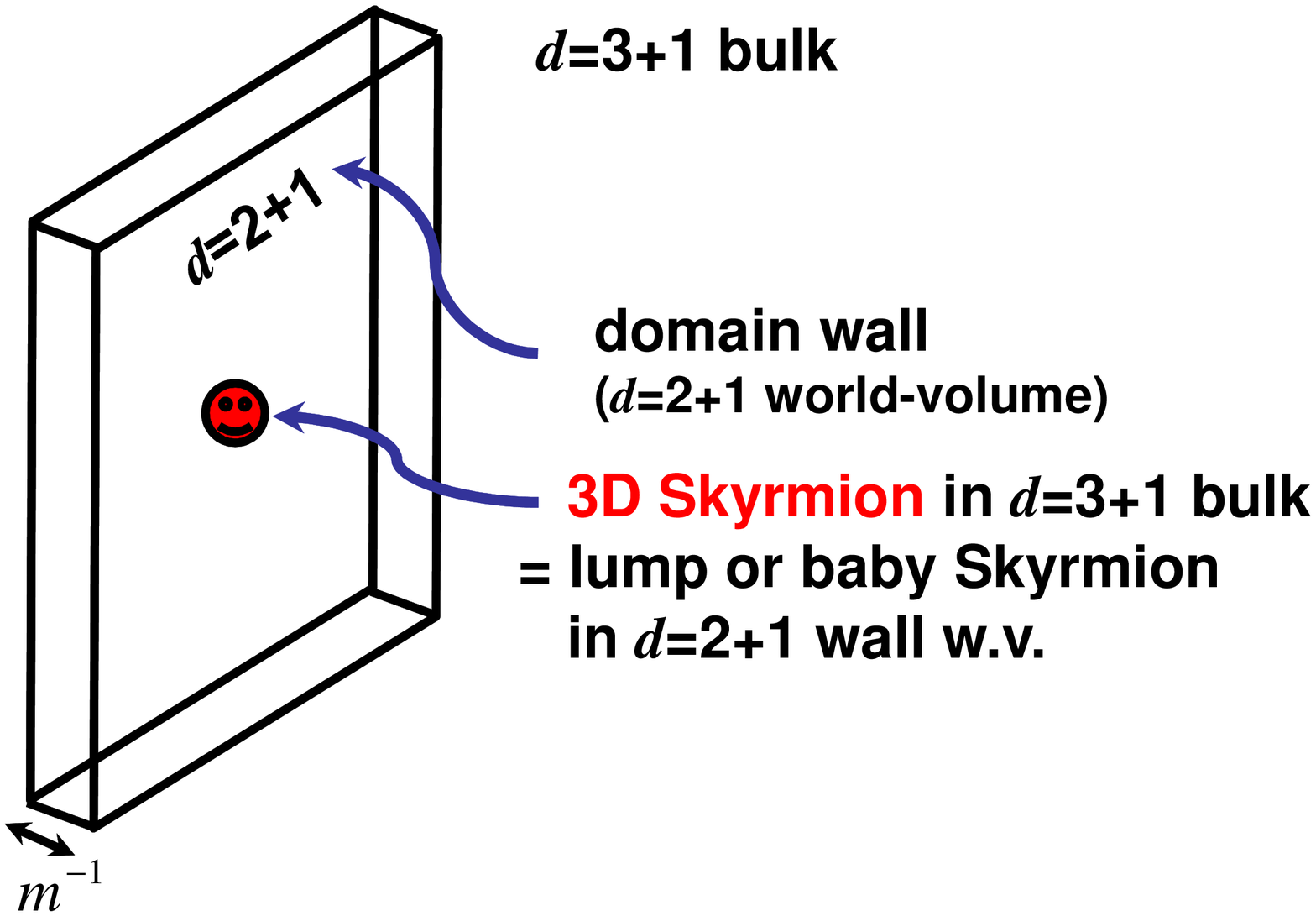}&
\includegraphics[width=0.5\linewidth,keepaspectratio]{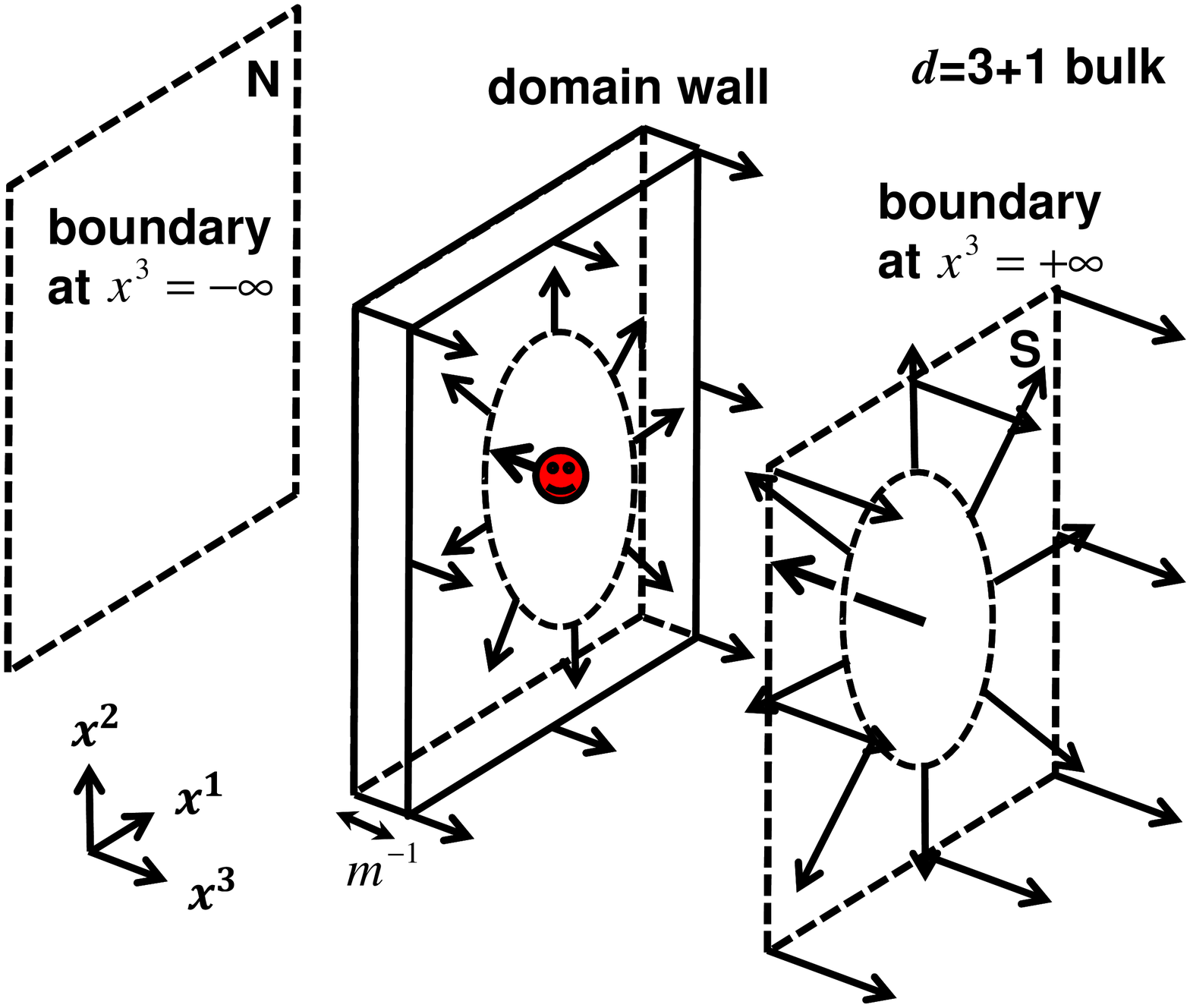}
\\
(a) & (b)
\end{tabular}
\caption{(a) A lump inside a domain wall corresponding to a 3D Skyrmion in the bulk.
(b) Texture structure of a 3D Skyrmion trapped inside the domain wall. 
Arrows denote points in the target space $S^3$. 
The length of the arrows denotes the distance from the north pole of $S^3$: 
arrows with zero, medium (one-half), and the maximal length (one) denote the north pole, 
the points on the $S^2$ at the equator, and the south pole, respectively. 
Note that all arrows in the maximal length represent the same point S on  
 the target space $S^3$.
At each slice of $x^3$= constant  except for $x^3 = \pm \infty$, 
the arrows wrap the sphere $S^2$ at $n_4$ = constant in the target space. 
Along the $x^3$-axis from $x^3=-\infty$ to $x^3=+\infty$, 
the length of arrows continuously changes from zero to one.
\label{fig:skyrmion}}
\end{center}
\end{figure}
What do they correspond to in the $3+1$-dimensional bulk?
They are nothing but 3D Skyrmions characterized by $\pi_3(S^3) \simeq {\bf Z}$; 
see Fig.~\ref{fig:skyrmion}(a).
In order to see this correspondence, 
we schematically plot the spin texture of a $k=1$ lump solution 
in Fig.~\ref{fig:skyrmion}(b).
Here, the arrows denote points in the target space $S^3$.
The length of the arrows denotes the distance from the north pole of $S^3$: 
arrows with the lengths zero, medium ($\pi/2$), and the maximal length ($\pi$) 
denote the north pole N, 
the points on the $S^2$ at the equator, and the south pole S, respectively. 
At each slice of $x^3$= constant, the arrows wrap the sphere $S^2$ at $n_4$ = constant 
in the target space. 
Along the $x^3$-axis from $x^3=-\infty$ to $x^3=+\infty$, 
the length of arrows continuously changes from zero to $\pi$.
Then, comparing this with Fig.~\ref{fig:3sphere}, 
one can find that the total configuration wraps the whole $S^3$ once 
so that it corresponds to the unit element of  $\pi_3(S^3) \simeq {\bf Z}$.

We have a few remarks.
In the absence of the Skyrme term, 
3D Skyrmions are unstable to shrink in the bulk, 
as is well known from the Derrick's scaling argument \cite{Derrick:1964ww}. 
However, what we have found here is that they can stably exist inside 
a domain wall! 
Moreover, as can be seen from multi-lump solutions, 
they can be placed at any positions in the $x^1$-$x^2$ plane 
in the domain wall world-volume 
which implies that there exists no static force between multiple Skyrmions.
They further have size and $U(1)$ phase moduli.  
This situation is very different from the original Skyrmions. 
In a sense, this is an example of an analytic solution of the 3D Skyrmion solution.

From the construction in the last subsection, 
we see that the non-Abelian domain wall is BPS. 
Moreover, the lump solution in the effective theory is also BPS. 
These two facts imply that 3D Skyrmions as this composite state are also BPS. 
So far, we do not know whether the theory can be embedded into a 
supersymmetric theory in which these solitons preserve a fraction of supercharges.

Similar situation occurs for Yang-Mills instantons (particles) 
in gauge theories coupled with Higgs fields in $d=4+1$ dimensions.
Yang-Mills instantons shrink to zero size (small instantons) 
in the presence of the Higgs coupling, 
while they can stably exist as lumps \cite{Eto:2004rz} 
inside a non-Abelian vortex \cite{Auzzi:2003fs,Eto:2006pg} . 
Both the non-Abelian vortices and lumps are BPS, 
and thus this composite state is also BPS.
In this case, the theory can be embedded into 
supersymmetric gauge theories with eight supercharges, 
in which non-Abelian vortices and lumps are 1/2 BPS states 
preserving half of the supersymmetry and
the composite state is a 1/4 BPS state preserving a quarter of supersymmetry \cite{Eto:2004rz}.
We summarize in Table~\ref{table:BPSlumps}
the two cases of 3D Skyrmions and Yang-Mills instantons 
realized as BPS lumps in the non-Abelian domain wall and vortex, respectively.
\begin{table}[ht]
\begin{tabular}{c|ccccc} \hline
&  $d=2+1$ &                         & $d=3+1$           & $d=4+1$  \\ \hline
this paper & Lumps     & $\longleftarrow$ & 3D Skyrmions   &  \\ 
           &    &  Non-Abelian  wall                          &  \\ \hline
 \cite{Eto:2004rz} & Lumps   & $\longleftarrow$ & $\longleftarrow$  & Yang-Mills instantons \\
               &                     &     & Non-Abelian vortex                \\ \hline                 
\end{tabular}
\caption{\label{table:BPSlumps}
BPS 3D Skyrmions and Yang-Mills instantons 
as lumps in the non-Abelian domain wall and vortex, respectively.
Both BPS 3D Skyrmions and Yang-Mills instantons cannot stably 
exist in the bulk where they are unstable to shrink. 
They can stably exist inside 
the non-Abelian domain wall and vortex, respectively; 
they are realized as BPS lumps in the world-volume theory of the 
 non-Abelian domain wall and vortex.
}
\end{table}

Yang-Mills instantons can exist 
also inside a non-Abelian domain wall \cite{Eto:2005cc}. 
Yang-Mills instantons become 3D Skyrmions 
in the effective theory of the domain wall, 
which gives a physical realization of the 
Atiyah-Manton ansatz of 3D Skyrmion from 
instanton holonomy \cite{Atiyah:1989dq}.
However, in this case, 
the composite state is not BPS although 
the instantons and the domain wall are both BPS. 

\subsection{Spinning Skyrmions inside the wall as Q-lumps}
In the presence of the additional potential term 
of either Eq.~(\ref{eq:pot2}) or (\ref{eq:pot3}), 
lumps are unstable to shrink. 
Instead, there exist stable Q-lumps \cite{Leese:1991hr}, 
time-dependent stationary lump solutions,  
which may correspond to spinning Skyrmions in 
the $d=3+1$-dimensional bulk. 
Here, we consider the mass of type in Eq.~(\ref{eq:pot3}) with $\hat i=3$.
In addition to the static energy in Eq.~(\ref{eq:BPSeq-lump}), 
there are contributions from the time dependence and the mass: 
\beq
 {2m \over \sqrt 2}
 E_{\rm time+ mass} &=& \int d^2 x 
 {\partial_0 u^* \partial_0 u 
 + M^2 |u|^2
  \over (1+|u|^2)^2} \non
 &=& \int d^2 x \left[
  {|\partial_0 u \mp i M u|^2 \over (1+|u|^2)^2} 
   \pm {i M (\partial_0 u^* \cdot u - u^* \partial_0 u )
     \over  (1+|u|^2)^2}
  \right] \non
 &\geq& \left| \int d^2 x \,
  {i M (\partial_0 u^* \cdot u - u^* \partial_0 u )
     \over  (1+|u|^2)^2} 
 \right| = |Q_{\rm lump}|,
\eeq
with the induced mass $M \equiv \sqrt 2 m_3$ and the Noether charge defined by 
\beq
Q_{\rm lump} \equiv
\int d^2 x \,
  {i M (\partial_0 u^* \cdot u - u^* \partial_0 u )
     \over  (1+|u|^2)^2}  .
\eeq
Then, the total energy is bound from below as 
$ {2m \over \sqrt 2}E_{\rm total} =  {2m \over \sqrt 2}(E + E_{\rm mass}) \geq |T| + |Q|$.
We have the BPS equation 
\beq
\partial_0 u \mp i M u =0,
\eeq
for time dependence in addition to the BPS lump equation (\ref{eq:BPS-lump-eq}).
We immediately obtain  
\beq 
 u(z,t) = u_{\rm lump}(z) e^{\pm i M x^0},
\eeq
with the lump solution $u_{\rm lump}(z)$ in Eq.~(\ref{eq:lump-sol}).
As denoted above, this solution may correspond 
to spinning Skyrmions in the $d=3+1$ bulk.

\subsection{Conventional 3D Skyrmions inside the wall as baby Skyrmions: 
With the Skyrme term}
BPS 3D Skyrmions studied in the last subsection are unstable in the bulk. 
In order to stabilize 3D Skyrmions in the bulk, we need to add the the Skyrme term (\ref{eq:Skyrme-term}) in the original Lagrangian.
Here, we consider the effect of the Skyrme term (\ref{eq:Skyrme-term}) 
in the domain wall effective action. 

We turn on the Skyrme term ${\cal L}^{(4)}$ perturbatively. 
We thus obtain the following term in the domain wall effective action 
by performing the integration:  
\beq
   {\cal L}_{\rm dw.eff.} ^{(4)} 
&=&  \1{2}  \left(\int_{-\infty}^{+\infty} dx \sin^4 \theta \right)
 \left[
              (\partial_a\hat{\bf n} \cdot \partial_b\hat{\bf n})
              (\partial^a\hat{\bf n} \cdot \partial^b\hat{\bf n}) 
           - (\partial_a\hat{\bf n} \cdot \partial^a\hat{\bf n})^2
         \right]\non
&=&  {\sqrt 2 \over 3m} \left[
              (\partial_a\hat{\bf n} \cdot \partial_b\hat{\bf n})
              (\partial^a\hat{\bf n} \cdot \partial^b\hat{\bf n}) 
           - (\partial_a\hat{\bf n} \cdot \partial^a\hat{\bf n})^2
         \right]\non
&=& - {\sqrt 2 \over 3m} \left[
              (\partial_a\hat{\bf n} \times \partial_b\hat{\bf n})
              (\partial^a\hat{\bf n} \times \partial^b\hat{\bf n}) 
         \right] ,\label{eq:baby-skyrme}
\eeq
where we have used the third formula in Eq.~(\ref{eq:formulas}).
This is precisely the baby Skyrme term including the signature \cite{Piette:1994ug}.

If we consider only this term without the potential, the 2D Skyrmion will be unstable to expand. 
In order to stabilize it, we consider the potential term (\ref{eq:potential-n2}) 
in the bulk Lagrangian, which induces 
the potential Eq.~(\ref{eq:pot2}) or (\ref{eq:pot3}) in the domain wall effective action.
In the literature, these potential terms are called the old \cite{Piette:1994ug} 
and new \cite{Weidig:1998ii} baby Skyrme terms, respectively   
 \cite{Harland:2007pb}.
Thus, there exist stable 2D Skyrmions with a fixed size, 
so-called baby Skyrmions \cite{Piette:1994ug},  on the domain wall.
Since the topological charges are unchanged from the last subsection without 
the Skyrme term, 
these baby Skyrmions correspond to conventional 3D Skyrmions in the bulk 3+1 dimensions. 
The presence of the mass term does not change the stability of 3D Skyrmions in the bulk.
We summarize, in Table~\ref{table:3D2DSkyrmions}, 
the stability of 3D Skyrmions in the bulk and 
2D Skyrmions inside the wall.

\begin{table}[ht]
\begin{tabular}{|c|c|c|}
\hline
Terms $\backslash$ dimension & 2D Skyrmion in $d=2+1$ & 3D Skyrmion in $d=3+1$ \\ \hline
                 Non  & Marginally stable (BPS) & Unstable to shrink \\ \hline
Skyrme term      & Unstable to expand & Stable \\ \hline
Mass term          & Unstable to shrink & Stable \\ \hline
Skyrme + mass terms & Stable (baby) & Stable \\ \hline 
\end{tabular}
\caption{\label{table:3D2DSkyrmions}
Correspondence between (in)stabilities of 2D and 3D Skyrmions.
Here, ``stable" implies that it has a fixed size while 
``marginally stable" implies that its size can be changed with the same energy 
so that it has a size modulus. ``baby" implies baby Skyrmions.
}
\end{table}

Since 3D Skyrmions are stable in the bulk,  3D Skyrmions trapped inside 
the domain wall are allowed to leave from the domain wall.
However, it was shown in  \cite{Kudryavtsev:1999zm} that 
there exists an attraction between the 3D Skyrmions in the bulk and the domain wall.
Therefore the 3D Skyrmions are absorbed into the domain wall  
becoming the baby Skyrmions. 

\section{Summary and Discussion \label{sec:summary} }
We have clarified a relation between Skyrmions in 2+1 and 3+1 dimensions.
3D Skyrmions can be constructed as 2D Skyrmions 
in the non-Abelian domain wall with  the $S^2$ moduli.
In the absence of the Skyrme term in the bulk, 
3D Skyrmions realized as 2D Skyrmions (lumps)  on the wall are BPS, 
and no force exists among them. 
In this case,  BPS 3D Skyrmions can exist stably  inside the domain wall while 
they  are unstable against shrinkage in the bulk.
We have also constructed Q-lumps in the wall theory which can be understood as
spinning Skyrmions in the bulk point of view 
in the model with an additional mass term.
We then have shown that baby Skyrmions in the wall theory can be identified with 
conventional 3D Skyrmions in the bulk in the Skyrme model 
with a modified mass. 
The correspondence between 2D and 3D Skyrmions has been summarized in Table~\ref{table:3D2DSkyrmions}.

Some discussions are addressed here.
Without the Skyrme term in the original theory, we have realized BPS 3D Skyrmions 
as BPS lumps  on the BPS domain wall.
In general, BPS solitons can be naturally embedded into supersymmetric field theories 
in which they break and preserve a fraction of supersymmetry. 
Thus far, it is not clear if our theory can be embedded into supersymmetric theories. 
The minimum number of the supercharges in $d=4+1$ is eight, 
which requires the target space of sigma models to be hyper-K\"ahler 
\cite{AlvarezGaume:1981hm}. 
The target space of the $O(4)$ model is $S^3$ which is not hyper-K\"ahler, 
and therefore we should embed $S^3$ into a larger hyper-K\"ahler manifold.
Composite BPS states in supersymmetric theories with eight supercharges 
were classified in \cite{Eto:2005sw}. 
Realizing BPS 3D Skyrmions as possibly 1/4 BPS states remains as 
a  future problem.

If one introduces the potential term 
of the form  $V = m^2(1-n_4^2) + m_3^2 n_3^2$ 
with the plus sign for the second term, 
instead of the minus sign studied in this paper, 
a 2D Skyrmion (lump) can be split into 
a set of two merons (fractional vortices) carrying fractional charges \cite{Nitta:2011um}.
Therefore, after a 3D Skyrmion is absorbed into the domain wall, 
it is split into two pieces 
each of which carries a fraction of 3D Skyrme charge.
Such a splitting was studied in a lower-dimensional case \cite{Auzzi:2006ju} 
in which a lump absorbed into a domain wall is split into two 
sine-Gordon kinks. 

If we consider the Skyrme model with the modified mass (without additional mass terms) 
in $d=4+1$, we have a non-Abelian domain wall with the $S^2$ moduli, 
whose effective theory is precisely the Faddeev-Skyrme model 
(without any potential) in $d=3+1$ \cite{Faddeev:1996zj}. 
This model is known to admit a knot soliton characterized by 
the Hopf charge $\pi_3(S^2)\simeq {\bf Z}$. Then a question arises. 
What does a knot soliton correspond to in the bulk? 
However, in the bulk, the target space is $S^3$ and the space is four dimensional, 
so that the homotopy group tells $\pi_4(S^3)\simeq {\bf Z}_2 \neq  {\bf Z}$. 
This is a twisted closed 3D Skyrmion string \cite{Nakahara:1986kn}.
It may imply that only the parity of the Hopf charge remains in the bulk, 
when the knot soliton leaves from the domain wall. 
It is of course an open question if such a soliton is stabilized only by 
the Skyrme term. 

\section*{Acknowledgements}

This work is supported in part by 
Grant-in-Aid for Scientific Research (No. 23740198) 
and by the ``Topological Quantum Phenomena'' 
Grant-in-Aid for Scientific Research 
on Innovative Areas (No. 23103515)  
from the Ministry of Education, Culture, Sports, Science and Technology 
(MEXT) of Japan.


\newcommand{\J}[4]{{\sl #1} {\bf #2} (#3) #4}
\newcommand{\andJ}[3]{{\bf #1} (#2) #3}
\newcommand{\AP}{Ann.\ Phys.\ (N.Y.)}
\newcommand{\MPL}{Mod.\ Phys.\ Lett.}
\newcommand{\NP}{Nucl.\ Phys.}
\newcommand{\PL}{Phys.\ Lett.}
\newcommand{\PR}{ Phys.\ Rev.}
\newcommand{\PRL}{Phys.\ Rev.\ Lett.}
\newcommand{\PTP}{Prog.\ Theor.\ Phys.}
\newcommand{\hep}[1]{{\tt hep-th/{#1}}}

\end{document}